\documentclass[confrence]{IEEEtran}

\usepackage{amssymb}
\usepackage{amsmath}
\usepackage{graphicx}
\usepackage{setspace}
\usepackage{euscript}
\usepackage{balance}
\usepackage{cite}

\begin{document}
%
\title{Hadamard Coded Modulation:\\An Alternative to OFDM for Optical Wireless Communications}

\author{\IEEEauthorblockN{Mohammad Noshad and Ma\"{\i}t\'{e} Brandt-Pearce}\\
\IEEEauthorblockA{Charles L. Brown Department of Electrical and Computer Engineering\\
University of Virginia\\ Charlottesville, VA 22904\\
Email: noshad@virginia.edu, mb-p@virginia.edu}
}

\markboth{}{Shell \MakeLowercase{\textit{et al.}}: Bare Demo of IEEEtran.cls for Journals}

\maketitle
\thispagestyle{empty}
\pagestyle{empty}

\begin{abstract}
Orthogonal frequency division multiplexing (OFDM) is a modulation technique susceptible to source, channel and amplifier nonlinearities because of its high peak-to-average ratio (PAPR).
The distortion gets worse by increasing the average power of the OFDM signals since larger portion of the signals are affected by nonlinearity.
In this paper we introduce Hadamard coded modulation (HCM) that uses the fast Walsh-Hadamard transform (FWHT) to modulate data as an alternative technique to OFDM in direct-detection wireless optical systems. This technique is shown to have a better performance for high average optical power scenarios because of its small PAPR, and can be used instead of OFDM in two scenarios: 1) in optical systems that require high average optical powers such as visible light communications (VLC), and 2) in optical wireless systems unconstrained by average power, for which HCM achieves lower bit error rate (BER) compared to OFDM. The power efficiency of HCM can be improved by removing a part of the signal's DC bias without losing any information. In this way, the amplitude of the transmitted signal is decreased and the signals become less susceptible to nonlinearity. Interleaving can be applied on HCM to make the resulting signals resistent against inter-symbol interference (ISI) effects in dispersive channels by uniformly distributing the interference over all symbols.

\end{abstract}

\begin{keywords}
Wireless optical communications, Walsh-Hadamard transform, orthogonal frequency division multiplexing (OFDM), peak-to-average power ratio (PAPR), nonlinear systems.
\end{keywords}

\section{Introduction}  \label{sec: Introduction}

Orthogonal frequency-division multiplexing (OFDM) is a high-dimensional modulation technique for high data-rate transmission that has been widely adapted to many modern broadband communications and standards because of its high spectral efficiency, simple one-tap frequency domain equalization, and robustness against narrow-band interference \cite{OFDM-Book}.
One main problem of OFDM is that the transmitted signals have a high peak-to-average power ratio (PAPR), which imposes serious signal distortion at the output of nonlinear channels \cite{OFDM-PAPR-New-Direction-13}. In this paper we propose an alternative to OFDM based on the Hadamard transform that maintains many of the advantages of OFDM but is notably more resilient against nonlinearity.

OFDM is also being considered as a modulation technique for energy efficient optical communications because of its better average-power efficiency compared to other spectrally efficient schemes \cite{Optical-OFDM-Armstrong-13}.
Modified forms of OFDM have been introduced \cite{Optical-OFDM-Armstrong-09} to make the application of OFDM possible to intensity-modulation direct-detection (IM/DD) optical communication systems. These types of OFDM generate non-negative real signals by modulating only some of the subcarriers and applying Hermitian symmetry on the data. Similar to the original OFDM, these techniques generate signals with large peaks, which is a big drawback in optical communication systems. The optical sources used in wireless optical communications (WOC) have a peak-power limit, forcing the output optical signals to have limited peak amplitudes.
This peak-power constraint causes a distortion on the OFDM signals. Increasing the transmitted average power makes the signals more likely to be clipped, and consequently, the distortion of the optical signals generated becomes larger. Therefore, the performance of OFDM is limited by distortion in systems that require high average optical powers.
Visible light communications (VLC) is an example of WOC that has high average optical power requirement since it is supposed to also fulfill the lighting needs, and the illumination is proportional to the average optical power.

Many modulation schemes are proposed to address the distortion effect in peak-power limited systems. Pulsed techniques are shown to have a good performance in IM/DD optical systems with peak-power constraints since they use the sources in their on or off mode \cite{EPPM12}. A problem with these modulation techniques is their low-spectral efficiency, which either limits the throughput or leads to high bit error rates (BER) in dispersive channels. A solution to this problem is increasing the number of signal amplitudes and sending multilevel symbols to increase the constellation size. The application of an array of sources instead of a single source gives the potential to generate multilevel signaling by separately turning on and off each source in the array \cite{Multilevel-EPPM12}. White LED bulbs that use 60-100 LEDs to provide illumination are examples of arrays that have been proposed for use in VLC systems. The increased constellation cardinality in the technique proposed in \cite{Multilevel-EPPM12} is achieved by trading-off low complexity.

Hadamard matrices and the Hadamard transform are popular tools in communication systems. They have been proposed as a precoder in OFDM systems to decrease the PAPR \cite{OFDM-Hadamard-11, OFDM-Hadamard-12}, reduce the BER \cite{OFDM-Hadamard-BER-03} and increase the resistance of the signals against frequency selective fading \cite{OFDM-Hadamard-Fading-10}. In this paper we use the Hadamard transform not as a precoder, but as a modulation technique to encode and transmit the information. We introduce a multilevel signaling technique with low complexity for application to OWC. The proposed technique, which is named Hadamard coded modulation (HCM), uses binary Hadamard matrices to encode the input data stream.
This technique can be implemented using the fast Walsh-Hadamard transform (FWHT), which has the same complexity as the FFT used in OFDM, $N \log_2 N$, where $N$ is the size of the Hadamard matrix. HCM is shown to have a PAPR of 2, which is significantly lower than that of OFDM, and can therefore provide brighter illumination levels in VLC systems compared to OFDM for the same BER. Because of its low PAPR, the waveforms of HCM are not clipped by the LEDs for average optical powers lower than half of the LED peak-power.

A version of the HCM waveforms is then proposed that has a reduced DC bias yet has the same performance as HCM but requires a lower average power, which is important in mobile applications. Based on simulation results, this technique can decrease the average optical power by half for $N=128$. Because of the lower amplitudes of the signals in this technique, the distortion due to clipping is smaller compared to the original HCM. In addition, applying interleaving on the HCM signals, as was proposed in \cite{VLC-JLT-I-13}, is shown to decrease the effect of interference on the HCM signals and lower the error probability in dispersive channels.

The rest of the paper is organized as follows. Section~\ref{sec:Problem Description} describes the principles of OFDM and its modified forms that are adapted to optical communications. In Section~\ref{sec:HCM} we introduce HCM and propose modified forms of HCM to increase its power efficiency and improve its performance in dispersive channels. Numerical results are presented in Section~\ref{sec:Numerical Results} that compare the performance of HCM to OFDM in dispersive and non-dispersive nonlinear channels. Finally, conclusions are drawn in Section~\ref{sec:Conclusion}.


\section{Problem Description} \label{sec:Problem Description}
This section describes OFDM and its modified forms proposed for IM/DD optical systems.
In this work we represent vectors with boldfaced lower-case letters, and boldfaced upper-case letters are reserved for matrices. The notation $\mathbf{A}^{\text{T}}$ denotes the transpose of the matrix $\mathbf{A}$, and $x^*$ indicates the conjugate of the complex number $x$. The notations $\mathbf{A}-x$ and $\mathbf{a}-x$ are respectively used to show the matrix and vector that are obtained by subtracting a scalar $x$ from all elements of the matrix $\mathbf{A}$ and vector $\mathbf{a}$.


\subsection{OFDM for Optical Communications}
\begin{figure} [!t]
    \begin{center}
    {\includegraphics[width=3.5in]{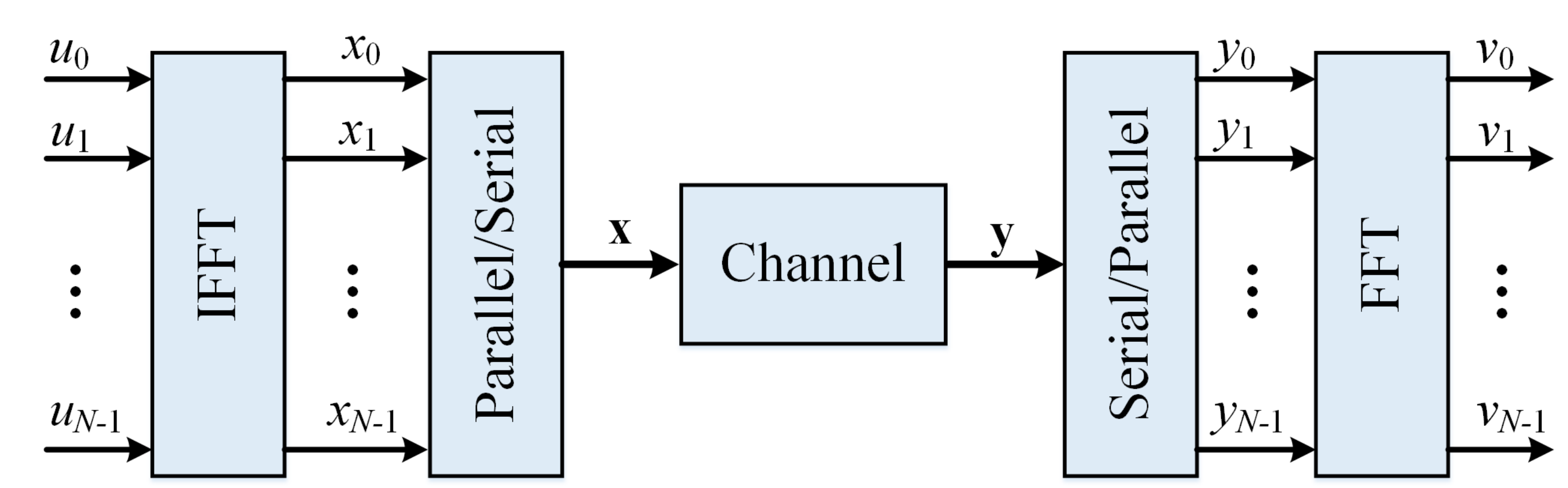}}
    \end{center}
    \caption{Block diagram of an OFDM system.}
    \label{OFDM}
    \end{figure}

Fig.~\ref{OFDM} shows a simple block-diagram of an OFDM system in which details such as adding a cyclic prefix are omitted for simplicity.
An OFDM symbol, $\mathbf{x}=[x_0, x_1, \cdots, x_{N-1}]$, is the IFFT of the data vector $\mathbf{u}=[u_0,u_1,\cdots,u_{N-1}]$, and is given by the $N$ point complex modulated sequence
    \begin{align}\label{IFFT}
        \mathbf{x} = \frac{1}{N} \mathbf{u} \mathbf{W},
    \end{align}
where $\mathbf{W}$ is an $N\times N$ matrix with $e^{ 2\pi j k n/N}$ as its $(n,k)$th element and $j^2=-1$. At the receiver side, an FFT is applied on the received signal to decode the data.

Asymmetrically clipped optical OFDM (ACO-OFDM) and DC-biased optical OFDM (DCO-OFDM) are the most popular modified forms of OFDM adapted for IM/DD optical communication systems \cite{ACO-OFDM-06, DCO-OFDM-06, DCO-OFDM-94}. In ACO-OFDM, $\mathbf{u}$ has instead the form
    \begin{align}\label{ACO-OFDM Input Data}
        \mathbf{u} = [0,u_0,0,u_1,\dots,u_{\frac{N}{4}-1}, 0 , u{^*_{\frac{N}{4}-1}},\dots,u{^*_1},0,u{_0^*}],
    \end{align}
in order to generate a real $\mathbf{x}$. Since $\mathbf{u}$ has Hermitian symmetry, the negative part of $\mathbf{x}$ can be clipped without losing any information.

DCO-OFDM uses both odd and even subcarriers as
    \begin{align}\label{DCO-OFDM Input Data}
        \mathbf{u} = [0,u_0,u_1,\dots,u_{\frac{N}{2}-2},0, u{^*_{\frac{N}{2}-2}},\dots,u{^*_1},u{_0^*}],
    \end{align}
and then a DC bias is added and whatever residual negative part is clipped. The modified OFDM current signals are then fed into an optical source to generate optical power proportional to that signal. In this paper we consider only ACO-OFDM, as it has been shown to be superior to DCO-OFDM \cite{Optical-OFDM-Armstrong-13}.

\begin{figure} [!b]
    \begin{center}
    {\includegraphics[width=2.0in]{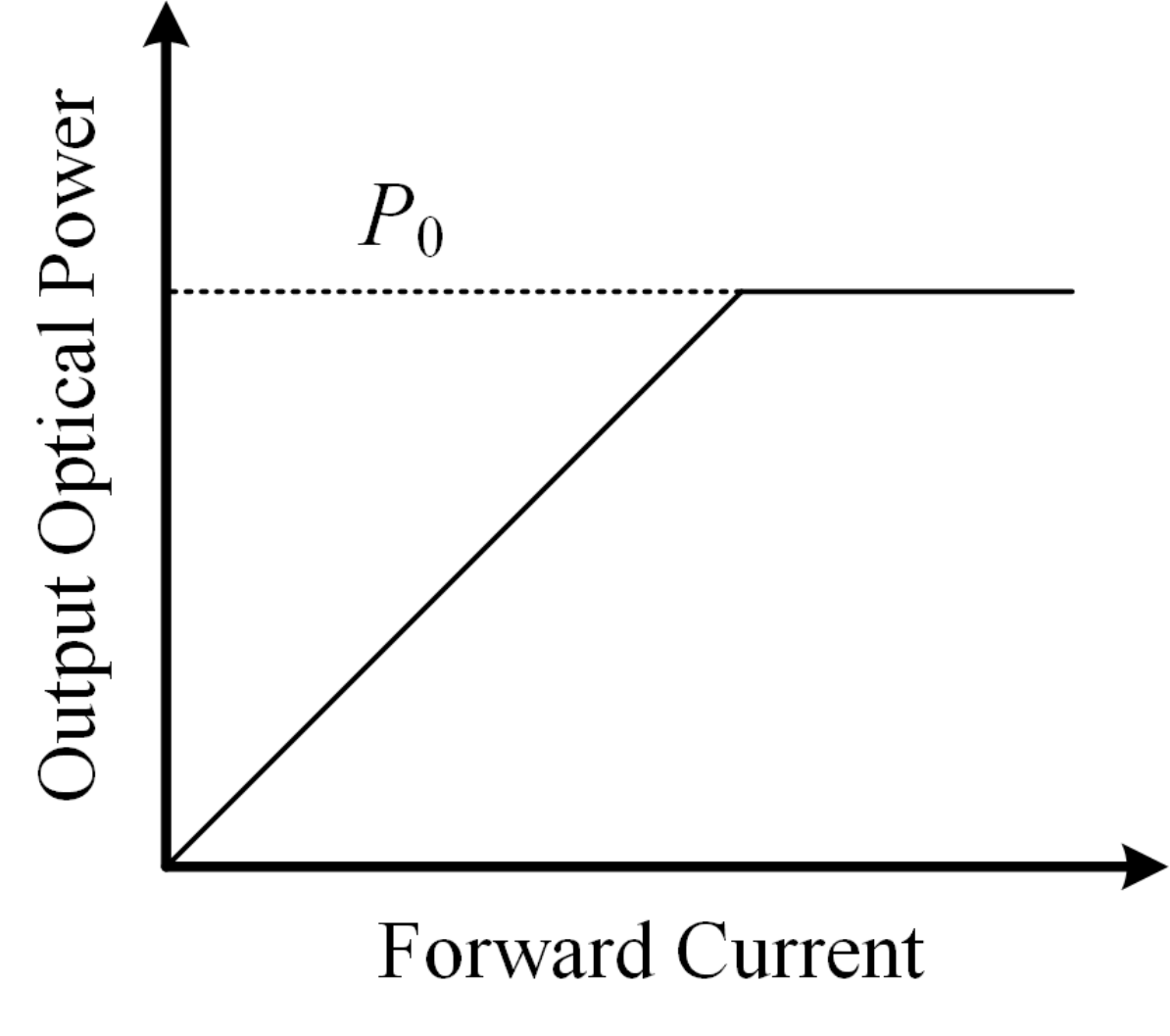}}
    \end{center}
    \caption{Output optical power of an ideal optical source with peak-power $P_0$ versus its forward input current.}
    \label{Ideal-LED}
    \end{figure}

As mentioned above, the sources in optical communication systems are peak-power limited, which clips the high peaks of the transmitted signal. In this paper we consider an ideal peak-power limited source, i.e., a hard limiter, which generates a power ranging from 0 to $P_0$ proportional to the forward input current (Fig.~\ref{Ideal-LED}), and the only distortion on the signals are assumed to be caused by clipping the transmitted signals below 0 and above $P_0$.

For large $N$, the OFDM signal can be modeled as a zero-mean Gaussian random variable with a variance of $\sigma^2$ \cite{OFDM-Book}. Hence, the probability distribution function of the light intensity in ACO-OFDM can be approximated as
    \begin{align}\label{}
        f_{\text{ACO}}(x) = \frac{1}{\sqrt{2 \pi}\sigma} &\exp\Big( -\frac{x^2}{2\sigma^2} \Big) \text{rect}\Big(\frac{x}{P_0}-\frac{1}{2}\Big) \nonumber\\& \hspace{0.5cm}+ Q\Big(\frac{P_0}{\sigma}\Big)\delta(x-P_0)+ \frac{1}{2}\delta(x),
    \end{align}
where $\delta(\cdot)$ is the Dirac delta function, $Q(\cdot)=\frac{1}{\sqrt{2\pi}} \int \exp\left( -x^2/2 \right)\, dx$ is the Gaussian $Q$-function and $\text{rect}(\cdot)$ is the rectangular function defined as
    \begin{equation}\label{}
        \text{rect}(x) =  \left\{ \begin{array}{ll}
                                        1 \;  & -\frac{1}{2} \le x \le \frac{1}{2},      \\
                                        0 \;          & \text{otherwise}
                                        \end{array}
              \right. .
    \end{equation}
Then the average optical power of ACO-OFDM can be obtained as
    \begin{align}\label{}
        P_{\text{ACO}} = \int f_{\text{ACO}}(x) dx =  \frac{\sigma}{\sqrt{2 \pi}} \Big( 1-e^{-\frac{P{_0^2}}{2\sigma^2}} \Big) + P_0 Q\Big(\frac{P_0}{\sigma}\Big).
    \end{align}

As described in \cite{Optical-OFDM-Clipping-Hass-11}, the distortion induced by the clipping can be modeled as a Gaussian noise with a variance of $\sigma{^2_{\textmd{lc}}}+\sigma{^2_{\textmd{uc}}}$, where $\sigma{^2_{\textmd{lc}}}$ and $\sigma{^2_{\textmd{uc}}}$ are the variances of the noises due to the clipping of the lower and and upper peaks of the OFDM signal, respectively. In ACO-OFDM, clipping the negative part of the signal causes a noise that is orthogonal to the data, and therefore does not affect the data, i.e., $\sigma{^2_{\textmd{lc}}}=0$. The variance of the clipping noise in ACO-OFDM can be obtained as \cite{Optical-OFDM-Clipping-Hass-11}
    \begin{align}\label{}
        \sigma{^2_{\textmd{uc}}} = (P{^2_0}+\sigma^2) Q\Big( \frac{P_0}{\sigma} \Big) - \frac{P_0 \sigma}{\sqrt{2\pi}} e^{-\frac{P{^2_0}}{2\sigma^2}}.
    \end{align}

The signal-to-noise ratio (SNR) at the receiver is given by \cite{Optical-OFDM-Clipping-Hass-11}
    \begin{align}\label{}
        \textmd{SNR} = \frac{\sigma^2}{\sigma{^2_n} + \sigma{^2_{\textmd{uc}}}},
    \end{align}
where $\sigma{^2_n}$ is the power of the discrete-time equivalent additive Gaussian noise. Assuming an $M$-QAM constellation with square $M$, the BER is given by \cite{Optical-OFDM-Clipping-Hass-11, BER-QAM-02}
    \begin{align}\label{}
        \textmd{BER}_{\textmd{OFDM}} = \frac{2(\sqrt{M}-1)}{\sqrt{M}\log_2 \sqrt{M}} Q\Bigg( \sqrt{\frac{3\textmd{SNR}}{M-1}} \Bigg) .
    \end{align}

\section{Hadamard Coded Modulation} \label{sec:HCM}
In this section we introduce the Hadamard coded modulation (HCM) technique as an alternative to OFDM. This modulation uses a binary Hadamard matrix to modulate the input data.

Let $\mathbf{H}_N$ be the binary Hadamard matrix of order $N$, which is obtained by replacing -1 by 0 in the original $\{-1,1\}$  Hadamard matrix \cite{Hadamard-Book}. 
%
    \begin{figure} [!t]
    \begin{center}
    {\includegraphics[width=3.4in]{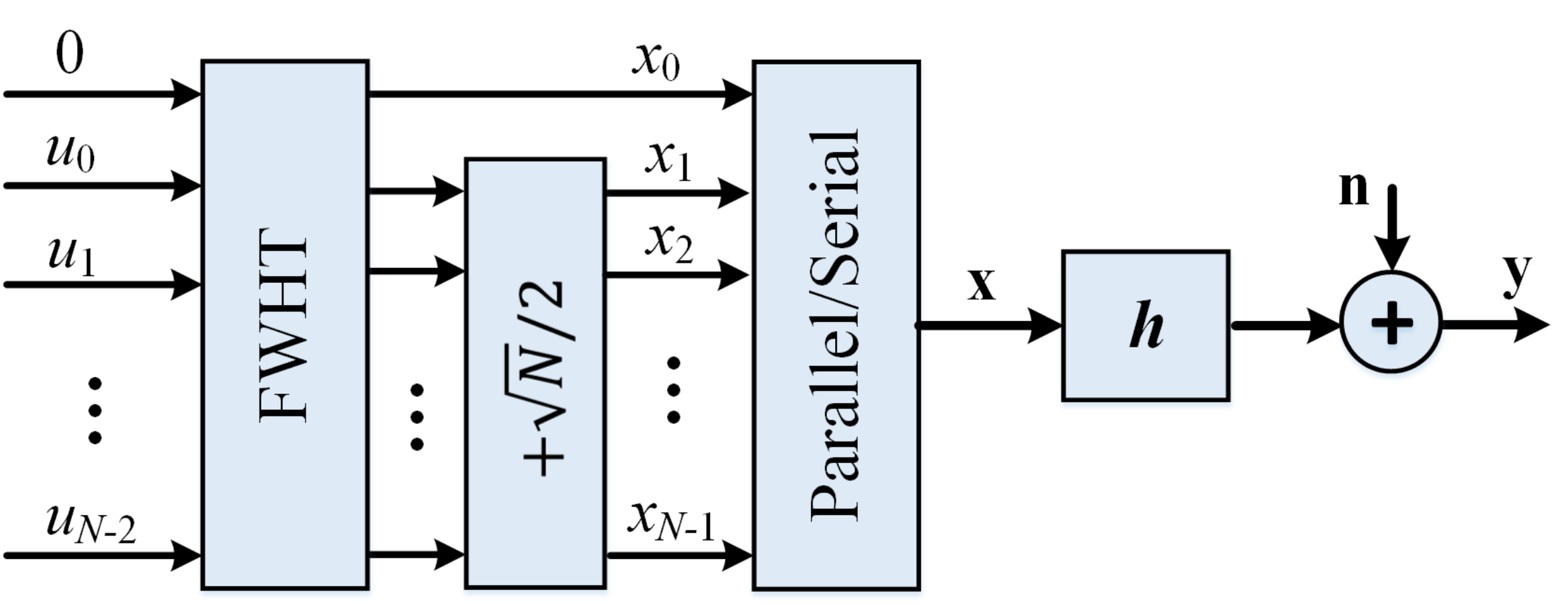}}
    \end{center}
    \caption{Block diagram of the HCM transmitter using FWHT.}
    \label{HCM-Transmitter}
    \end{figure}
    \begin{figure} [!b]
    \begin{center}
    {\includegraphics[width=2.2in]{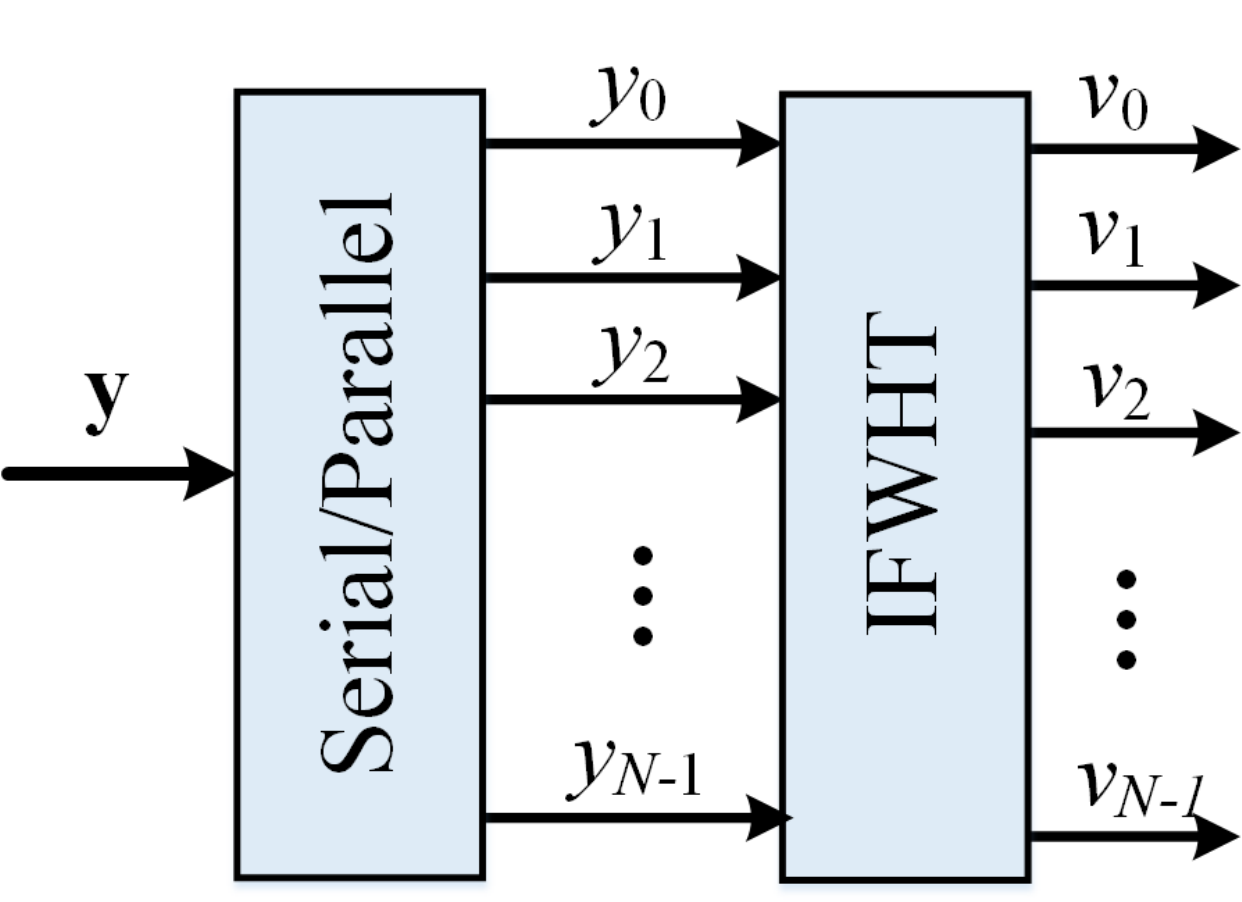}}
    \end{center}
    \caption{Block diagram of the HCM receiver using IFWHT.}
    \label{HCM-Receiver}
    \end{figure}
The components of $\mathbf{u}$ are assumed to be modulated using $M$-ary pulse amplitude modulation (PAM). The transmitted vector $\mathbf{x}$ is obtained from the data vector as
    \begin{align}\label{HCM Encoder}
        \mathbf{x} = \frac{1}{\sqrt{N}} \Bigg(\mathbf{u} \mathbf{H}{_N} + (1-\mathbf{u}) \overline{\mathbf{H}}{_N} \Bigg),
    \end{align}
where $\mathbf{\overline{H}}_N$ is the complement of $\mathbf{H}_N$. (\ref{HCM Encoder}) can be rewritten as
    \begin{align}\label{HCM Encoder using FWHT}
        \mathbf{x} =  \mathbf{u} \frac{1}{\sqrt{N}} \Big( \mathbf{H}{_N} - \overline{\mathbf{H}}{_N} \Big) - \frac{\sqrt{N}}{2}[0,1,1,\dots,1].
    \end{align}
where the second term is obtained from the product of a $1\times N$ vector of all ones and $\overline{\mathbf{H}}{_N}$. The matrix $( \mathbf{H}{_N} - \overline{\mathbf{H}}{_N})$ is the bipolar Hadamard matrix, and hence, the first term in (\ref{HCM Encoder using FWHT}) is the Walsh-Hadamard transform of the vector $\mathbf{u}$.
The transmitter can therefore be implemented using a fast Walsh-Hadamard transform (FWHT) as shown in Fig.~\ref{HCM-Transmitter}. The Hadamard transform is applied on the data stream using a FWHT of size $N$, which has a complexity of $N \log_2 N$, and then a constant value of $N/2$ is added to $N-1$ elements to generate $\mathbf{x}$.

In this work we only use $N-1$ rows of $\mathbf{H}{_N}$ that have a weight of $N/2$ to modulate the data, and we ignore the first row of the Hadamard matrix which has all ones. So, the first component of $\mathbf{u}$ is set to zero, and hence, the rate of $M$-PAM HCM is $(N-1)/N \log_2 M$. Then due to the fixed cross correlation between these remaining $N-1$ rows, the interference of the Hadamard codewords on each other can be removed at the receiver side \cite{EPPM12}, and the decoded vector $\mathbf{v}$ is obtained from the received vector $\mathbf{y}$  as
    \begin{align}\label{HCM Decoder}
        \mathbf{v} = \frac{1}{\sqrt{N}} \Big( \mathbf{y} \mathbf{H}{^\textmd{T}_N} - \mathbf{y} \mathbf{\overline{H}}{^\textmd{T}_N} \Big),
    \end{align}
which can be realized by an inverse FWHT (IFWHT) as shown in Fig.~\ref{HCM-Receiver}.

The noise due to the channel, $\mathbf{n}$, is assumed to be additive white Gaussian noise (AWGN), and the output signal is given by $\mathbf{y}=\mathbf{h}*\mathbf{x}+\mathbf{n}$, where $\mathbf{h}=\{h(k)\}$ is the discrete-time equivalent impulse response of the channel and $*$ denotes the convolution operation. For an ideal non-dispersive channel with impulse response
    \begin{equation}\label{}
        h(k) =  \left\{ \begin{array}{ll}
                                        1 \;  & k=0,      \\
                                        0 \;  & k\neq 0,
                                        \end{array}
              \right.
    \end{equation}
the decoded data can be rewritten as
    \begin{align}\label{Decoded Signal with Equivalent Noise}
        \mathbf{v} = \Big(\mathbf{u} - \frac{1}{2}[N-1,1,1\dots,1] \Big) + \mathbf{\tilde{n}},
    \end{align}
where $\mathbf{\tilde{n}} =  \frac{1}{\sqrt{N}} \mathbf{n} \Big( \mathbf{H}{^\textmd{T}_N} - \mathbf{\overline{H}}{^\textmd{T}_N}\Big)$ is a $1 \times N$ noise vector with independent components. In the derivation of (\ref{Decoded Signal with Equivalent Noise}) we have used the fact that
    \begin{equation}\label{}
        \mathbf{H}_N  \mathbf{H}{^\textmd{T}_N} =  \left[ \begin{array}{lllll}
                                        N           & N/2         & \dots & N/2         & N/2        \\
                                        N/2         & N/2         & \dots & N/4         & N/4         \\
                                        \vdots      & \vdots      &\ddots & \vdots      & \vdots      \\
                                        N/2         & N/4         & \dots & N/2         & N/4         \\
                                        N/2         & N/4         & \dots & N/4         & N/2
                                        \end{array}
              \right],
    \end{equation}
and
    \begin{equation}\label{}
        \mathbf{\overline{H}}_N  \mathbf{\overline{H}}{^\textmd{T}_N} =  \left[ \begin{array}{lllll}
                                        0           & 0           & \dots & 0           & 0           \\
                                        0           & N/2         & \dots & N/4         & N/4         \\
                                        \vdots      & \vdots      &\ddots & \vdots      & \vdots      \\
                                        0           & N/4         & \dots & N/2         & N/4         \\
                                        0           & N/4         & \dots & N/4         & N/2
                                        \end{array}
              \right].
    \end{equation}
%

For average power levels less than $P_0/2$, the HCM signals are not distorted by the LED nonlinearity and the performance of HCM is only limited by the noise. In this case, the optical average power is equal to $\sigma$, where $\sigma^2$ is the electrical HCM signal power. Through (\ref{Decoded Signal with Equivalent Noise}), the BER of $M$-PAM HCM can be calculated from \cite{BER-QAM-02} as
    \begin{align}\label{}
        \textmd{BER}_{\textmd{HCM}} = \frac{(M-1)}{M\log_2 M} Q\Bigg( \sqrt{\frac{3}{M^2-1}} \frac{\sigma}{\sigma_n} \Bigg) .
    \end{align}

\subsection{Reducing the DC Bias without Losing Information} \label{sec:Removing DC Bias}
It turns out that the signal $\mathbf{x}$ has an unnecessarily high mean value, which can be removed to improve the performance in nonlinear systems. In (\ref{HCM Decoder}), by adding a DC value $b_{\textmd{DC}}$ to the transmitted signal, $\mathbf{y}+b_{\textmd{DC}}$ is received instead of $\mathbf{y}$. Then, the decoded vector becomes
    \begin{align}\label{}
        \mathbf{v} = [\sqrt{N}b_{\textmd{DC}}, 0, \dots,0] + \mathbf{u} + \mathbf{\tilde{n}} - 1.
    \end{align}
Thus adding a DC level to the transmitted signal only affects the first component, and it has no effect on the rest of the data. As mentioned above, we are not sending information on the first component of $\mathbf{u}$, and therefore, we can remove a part of the DC component of the transmitted signal without losing any information. Hence, in order to increase the power efficiency, we preprocess the transmitted signal in order to set the minimum of each symbol to zero. This modified version of HCM is referred to as DC-removed HCM (DCR-HCM) in this paper. The transmitted signal of the DCR-HCM, $\widetilde{\mathbf{x}}$, is obtained from $\mathbf{x}$ as
    \begin{align}\label{}
        \widetilde{\mathbf{x}} = \mathbf{x} - \big(\min \mathbf{x} \big) ,
    \end{align}
and therefore, $\min \widetilde{\mathbf{x}}  = 0$.

\begin{figure} [!t]
   \vspace*{0 in}
    \begin{center}
    {\includegraphics[width=3.4in]{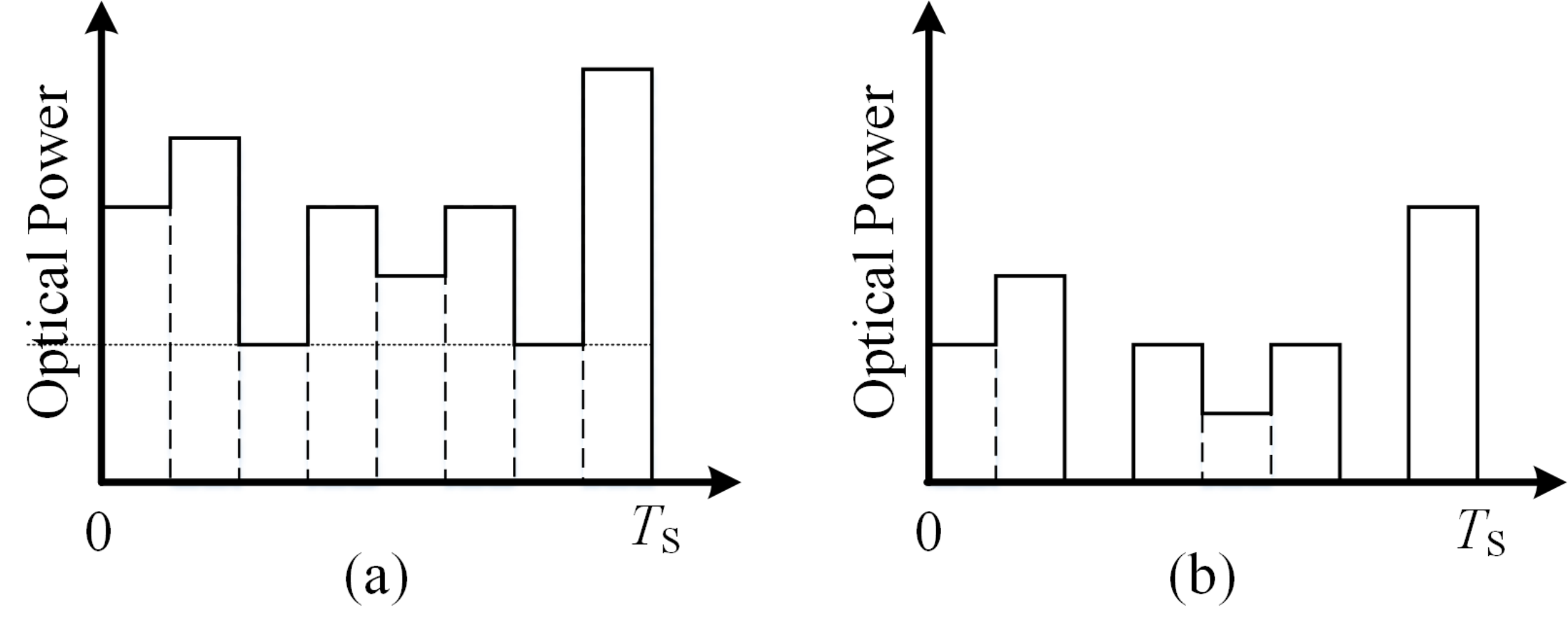}}
    \end{center}
    \vspace*{0 in}
    \caption{(a) An HCM signal, and (b) its corresponding DC removed signal.}
    \label{DC-Removal}
    \end{figure}

Fig.~\ref{DC-Removal} shows an example of DC removal in an HCM symbol, where $T_s$ is the symbol time, $\mathbf{x}$ is an HCM signal with $\min \mathbf{x}=2$, and its DC-removed form is obtained by setting $\widetilde{\mathbf{x}} = \mathbf{x} - 2$. As can be seen, the signals of DCR-HCM have lower amplitudes, making them less likely to be clipped by the optical source.

Through the following theorem we show that the average power of DCR-HCM is approximately half of that of HCM.

\vspace{0.1 in}
\emph{Theorem:} Let $\mathbf{x}=[x_0, x_1, \cdots, x_{N-1}]$ be an HCM waveform. Then,
    \begin{align}\label{}
        \max \mathbf{x} - \min \mathbf{x} \leq \frac{\sqrt{N}}{2}.
    \end{align}

\emph{Proof: } Let $\mathbf{x}$ be generated from the data vector $\mathbf{u}$, and $x_m$ and $x_n$ be respectively the smallest and largest elements of $\mathbf{x}$. Now let $\hat{\mathbf{x}}$ be the HCM waveform that has zero at its $m$th position and its other $N-1$ components are $\sqrt{N}/2$. Then we can get $\hat{\mathbf{x}}$ from $\mathbf{x}$ by changing only $x_m$ elements of $\mathbf{u}$ and forcing the $m$th component of $\mathbf{x}$ to be zero. By applying this change, $x_n$ in $\mathbf{x}$ becomes $\sqrt{N}/2$ in $\hat{\mathbf{x}}$. Hence, the difference between $x_n$ and $\sqrt{N}/2$ cannot be larger than $x_m$, which means that $x_n-\sqrt{N}/2 \leq x_m$.     $\hspace{5.5cm}\Box$
\vspace{0.1 in}

We know that if $\mathbf{x}_1$ is an HCM waveform generated by data sequence $\mathbf{u}_1$, then $\mathbf{x}_2=\frac{N-1}{\sqrt{N}}-\mathbf{x}_1$ is also an HCM waveform for the data vector $\overline{\mathbf{u}_1}$. As a result, $\min \mathbf{x}_2 = \frac{N-1}{\sqrt{N}}- \max \mathbf{x}_1$, and from the theorem above we get
    \begin{align}\label{Sum of Min in DCR-HCM}
        \min \mathbf{x}_1 + \min \mathbf{x}_2 \geq \frac{1}{\sqrt{N}}(N/2 -1).
    \end{align}
Through (\ref{Sum of Min in DCR-HCM}), at least a DC level of $\frac{N/2-1}{\sqrt{N}}$ is eliminated from any two complement signals, and therefore, DCR-HCM can decrease the average-power at least by a factor of
    \begin{align*}\label{}
        1-\frac{(N/2-1)N/2}{(N-1)N/2}=\frac{N}{2(N-1)},
    \end{align*}
which approaches 1/2 as $N$ increases.

\subsection{Interleaved HCM for Dispersive Channels} \label{sec:Interleaved HCM}
\begin{figure} [!t]
   \vspace*{0 in}
    \begin{center}
    {\includegraphics[width=3.4in]{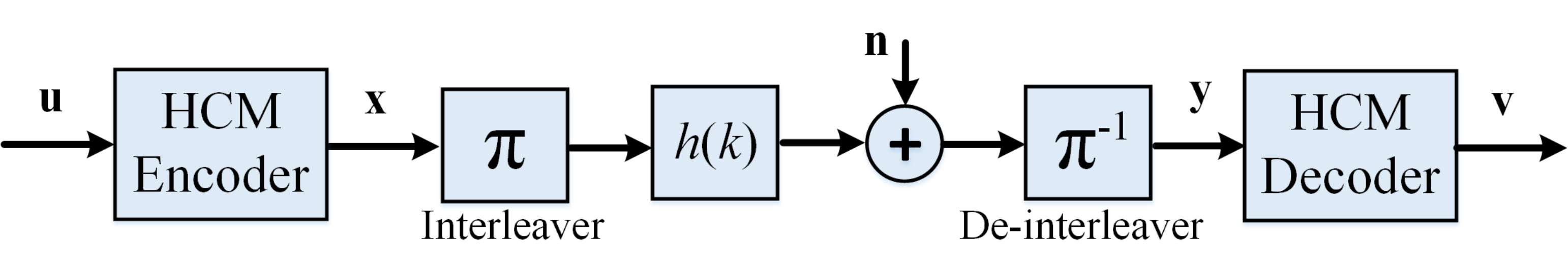}}
    \end{center}
    \vspace*{0 in}
    \caption{Schematic view of an interleaved HCM system for dispersive channels.}
    \label{Interleaved-HCM}
    \end{figure}

Optical wireless channels are dispersive channels that impose inter-symbol interference (ISI) on the transmitted signals, and therefore, any practical modulation technique must be renitent against ISI. In Hadamard matrices, some of the rows are cyclic shifts of one another, which makes HCM vulnerable to ISI. Interleaving has been shown to be an effective solution for this problem \cite{VLC-JLT-I-13}. In this technique, as shown in Fig.~\ref{Interleaved-HCM}, a symbol-length interleaver and deinterleaver are used at the transmitter and receiver to reduce the effects of intra-symbol ISI due to a dispersive channel with discrete impulse response $h(k)$. The interleaver is a permutation matrix, $\mathbf{\pi}$, and deinterleaver is its inverse, $\mathbf{\pi}^{-1}$. Hence, $\mathbf{x} \mathbf{\pi}$ is sent instead of $\mathbf{x}$. The best interleaver matrix is the one that evenly distributes the interference over all symbols, and can be found using binary linear programming (BLP) \cite{VLC-JLT-I-13}. In non-dispersive channels, the performance of interleaved HCM is the same as HCM since $\mathbf{\pi} \mathbf{\pi}^{-1}=1$. Similar to OFDM systems, adding a cyclic prefix decreases the interference between adjacent symbol interference.


%
    \begin{figure} [!t]
   \vspace*{0.1 in}
    \hspace{-0.0 in}{\includegraphics[width=3.4in]{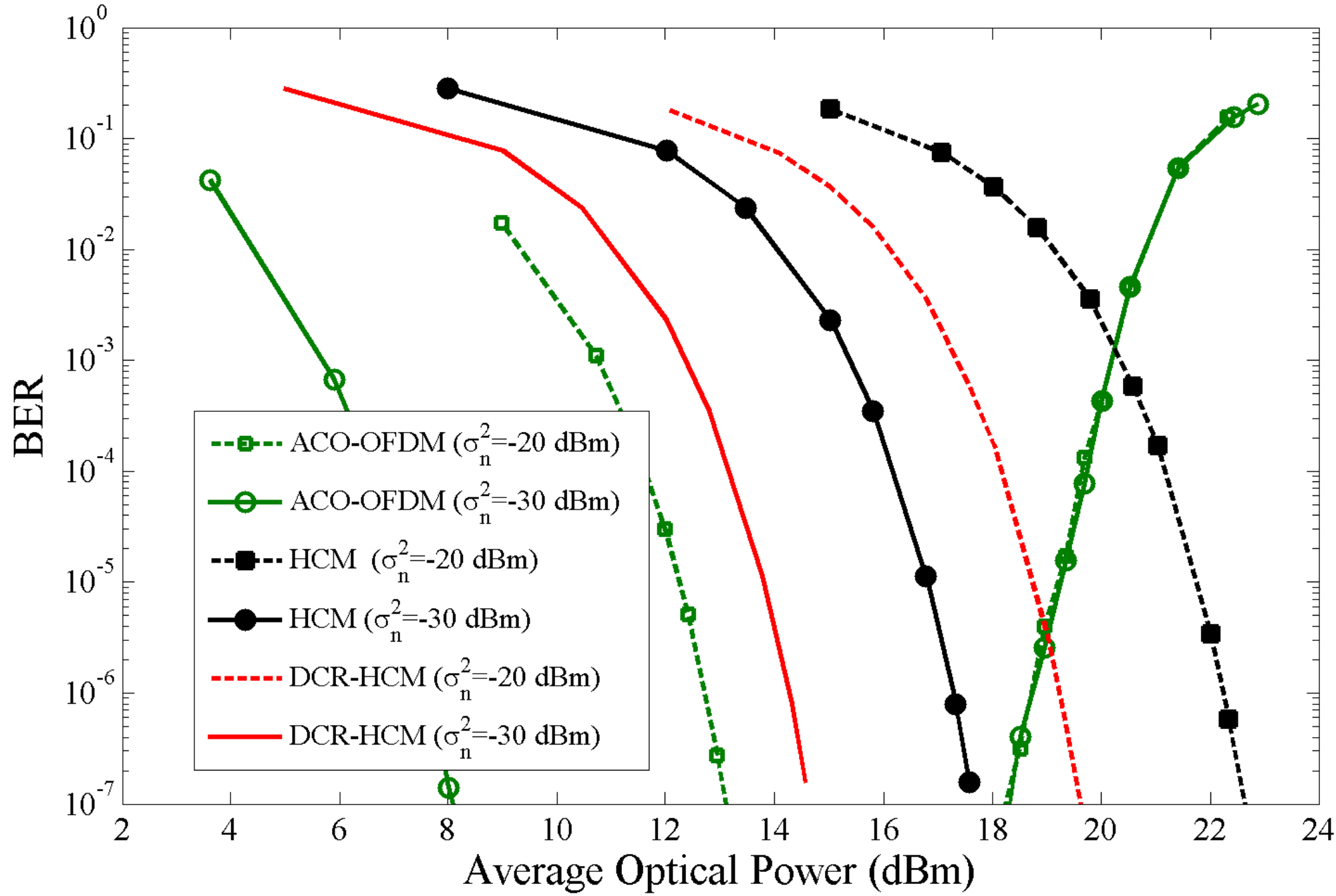}}
    \vspace*{0.1 in}
    \caption{BER of ACO-OFDM vs HCM and DCR-HCM in an ideal channel, i.e. h(0)=1, for two different noise power levels of $\sigma{^2_n}=-30$ dBm and $\sigma{^2_n}=-20$ dBm.}
    \label{BER-of-OFDM-vs-HCM}
    \end{figure}

\section{Numerical Results} \label{sec:Numerical Results}
In this section, numerical results using simulation 
are presented to compare the performance of HCM to ACO-OFDM. In the simulations, the sources are assumed to be ideal peak-power limited sources as shown in Fig.~\ref{Ideal-LED} with $P_0 = 0.5$ W, and perfect symbol synchronization is used. The data-rate is assumed to be 100 Mbps.
    \begin{figure} [!t]
    \hspace{-0.0 in}{\includegraphics[width=3.4in]{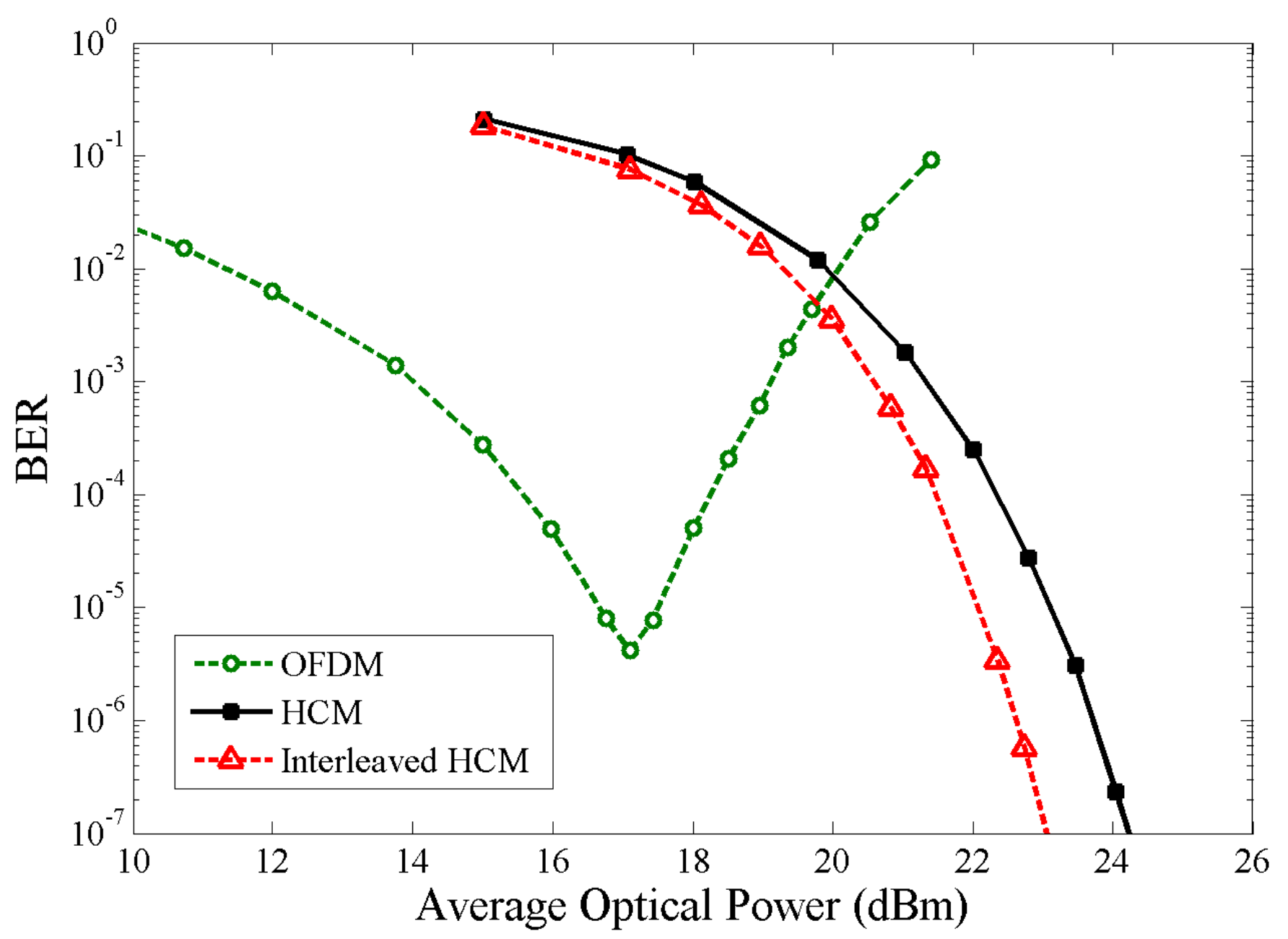}}
    \vspace*{0.1 in}
    \caption{BER of ACO-OFDM vs HCM and interleaved HCM in a dispersive channel with impulse response $h(0)=0.9$ and $h(1)=0.1$ and $\sigma{^2_n}=-20$ dBm noise power.}
    \label{BER-of-OFDM-vs-HCM-ISI}
    \end{figure}

The BER of ACO-OFDM, HCM and DCR-HCM is plotted versus the average received optical power in Fig.~\ref{BER-of-OFDM-vs-HCM}. The parameters of these techniques are chosen such that all have the same data-rate. The ACO-OFDM uses 16-QAM to modulate $N=128$ carriers, which has a spectral-efficiency of 1.\footnote{Note that simple on-off keying (with a spectral efficiency of 1 also) cannot be directly used in VLC systems because of light flicker.} The HCM signals are generated using an FWHT of size $N=128$, and on-off keying (OOK) is used to modulate the data. The results are plotted assuming a non-dispersive AWGN channel, for two noise levels of $\sigma{^2_n}=-30$ dBm and $\sigma{^2_n}=-20$ dBm. The BER of ACO-OFDM decreases by increasing the average optical power until it reaches a minimum value, and increases afterwards due to the clipping imposed distortion. As a result, HCM achieves lower BER for average optical powers higher than 18 dBm and 20.3 dBm for $\sigma{^2_n}=-30$ dBm and $\sigma{^2_n}=-20$ dBm, respectively. If the average power is increased beyond $P_0/2$ (24 dBm), HCM will begin to degrade as well due to the clipping. The power penalty experienced by HCM at low power levels is due to the unipolar codewords used to modulate the data. According to the simulation results in Fig.~\ref{BER-of-OFDM-vs-HCM}, the DCR-HCM uses 3dB lower average optical power than HCM for the same BER, which matches the results of Section~\ref{sec:Removing DC Bias}.

The performance of 16-QAM OFDM and OOK HCM are compared in Fig.~\ref{BER-of-OFDM-vs-HCM-ISI} for a dispersive channel with a discrete-time equivalent impulse response of $h(0)=0.9$ and $h(1)=0.1$ when no equalizer is used at the receiver. Both techniques use $N=128$ and a cyclic prefix of length 4. The noise is assumed to have a power of $\sigma{^2_n}=-20$ dBm. According to these results, HCM can achieve lower BERs compared to OFDM in this channel without considering any average-power constraint. The BER of interleaved HCM is also plotted versus the average optical power. The optimum interleaver is found using BLP as described in \cite{VLC-JLT-I-13}. As shown in Fig.~\ref{BER-of-OFDM-vs-HCM-ISI}, interleaving decreases the BER by alleviating the interference effect, approaching the energy performance of the non-dispersive case.

\section{Conclusion} \label{sec:Conclusion}
In this paper, HCM is introduced as an alternative technique to OFDM for application in 1) optical wireless communications requiring high average power levels, such as visible light communication (VLC) systems, and 2) systems with unconstrained average-power. HCM achieves a lower error probability floor compared to OFDM since it transmits a signal with substantially lower PAPR. DCR-HCM is introduced to decrease the average power required to achieve a given BER by removing a part of the DC bias from the HCM signals. Interleaving is shown to be an efficient technique to decrease the BER of HCM in ISI channels. In future work we plan to study the information capacity achievable with HCM  under various nonlinear system scenarios.

\bibliographystyle{IEEEtran}
\balance
\bibliography{EPPM,Thesis,OFDM}

\end{document}